\documentclass[fleqn,usenatbib]{mnras}

\usepackage{newtxtext,newtxmath}

\usepackage[T1]{fontenc}

\DeclareRobustCommand{\VAN}[3]{#2}
\let\VANthebibliography\thebibliography
\def\thebibliography{\DeclareRobustCommand{\VAN}[3]{##3}\VANthebibliography}

\usepackage{graphicx}	
\usepackage{amsmath}	
\usepackage{footnote}
\usepackage{tablefootnote}
\usepackage{physics}
\usepackage{caption}
\usepackage{subcaption}
\usepackage{graphicx}
\usepackage{mathcomp}
\usepackage{textcomp}

\usepackage{graphicx,import}

\usepackage{lipsum}

\usepackage{xcolor}

\title[Coupling starlight through turbulence into FMFs]{Starlight coupling through atmospheric turbulence into few-mode fibers and photonic lanterns in the presence of partial adaptive optics correction}

\author[M. Diab et al.]{
Momen Diab,$^{1}$\thanks{E-mail: mdiab@aip.de (AIP)}
Aline N. Dinkelaker,$^{1}$
John Davenport,$^{1}$
Kalaga Madhav$^{1}$\newauthor
and Martin M. Roth$^{1}$
\\
$^{1}$innoFSPEC, Leibniz Institute for Astrophysics Potsdam, An der Sternwarte 14482, Germany\\
}

\date{Accepted 2020 November 26. Received 2020 November 24; in original form 2020 October 30 }

\pubyear{2020}

\begin{document}
\label{firstpage}
\pagerange{\pageref{firstpage}--\pageref{lastpage}}
\maketitle

\begin{abstract}

Starlight corrupted by atmospheric turbulence cannot couple efficiently into astronomical instruments based on integrated optics as they require light of high spatial coherence to couple into their single-mode wave\-guides. Low-order adaptive optics in combination with photonic lanterns offer a practical approach to achieve efficient coupling into multiplexed astrophotonic devices. We investigate, aided by simulations and an experimental testbed, the trade-off between the degrees of freedom of the adaptive optics system and those of the input wave\-guide of an integrated optic component leading to a cost-effective hybrid system that achieves a signal-to-noise ratio higher than a standalone device fed by a single-mode fiber.

\end{abstract}










\section{Introduction}


As telescope apertures increase in diameter, optical instruments at their foci such as spectrographs need to proportionally expand in size to make use of the additional flux without compromising performance, e.g. resolving power or sensitivity~\citep{spano08, spano06}. 
This results in costly instruments with large physical dimensions, making them more sensitive to vibrational and environmental changes. 
Photonic technologies offer an opportunity to avoid bulk optics, thus limiting the increase in size. 
Using integrated optics (IO) to manipulate starlight in astronomical instruments before detection-- an emerging field known as astrophotonics-- has the potential of reducing the footprint and mass of astronomical instruments, cutting costs owing to simpler vacuum and thermal control, enhancing performance, and enabling multiplexing~\citep{minardi_astrophotonics_2020}.

Photonic spectrographs~\citep{blind17}, e.g. arrayed wave\-guide gratings (AWGs)~\citep{hawthorn06}, fiber Bragg gratings for OH suppression \citep{bland-hawthorn2011, rahman2020}, and photonic beam combiners, e.g. GRAVITY~\citep{eisenhauer_2008} and discrete beam combiners (DBCs)~\citep{minardi2015, minardi2012}, need to operate in the single mode regime in order to 
deliver their promised spectral resolution, filter characteristics and phase retrieval capabilities, respectively, while avoiding modal noise and focal ratio degradation. Coupling a seeing-limited point spread function (PSF) at the focus of a large telescope into a single mode wave\-guide is challenging and typically results in low efficiency. 
Two mitigation techniques can be applied to enable the use of a photonic instrument behind a ground-based telescope: On the one hand, an extreme adaptive optics (ExAO) system may be used to entirely correct for the atmospheric aberrations present in the received wavefronts, and in doing so convert the focal speckle pattern into a diffraction-limited spot that couples efficiently into a single-mode fiber (SMF). On the other hand, a photonic lantern can be employed to split the optical power coupled from the telescope into multiple SMFs~\citep{saval05}. ExAO systems have more degrees of freedom and run faster than conventional AO systems to deliver high Strehl ratios ($\mathit{SR} > 0.8$ in NIR) but can only do so for bright objects that act as their own guide stars~\citep{Guyon18}. As a result, they are more suited to high-contrast imaging of exoplanets and circumstellar disks. They also tend to be notoriously expensive for midsize ($2$ -- $4$ m) telescopes and can overwhelm the cost of the telescope itself. 

Photonic lanterns, conversely, are mode converting devices that redistribute multimodal light into multiple single-mode beams. They do so by guiding the light through an adiabatic taper from a multimode core to several single-mode cores. 
If the transition is gradual and the number of SMFs is equal to or greater than the number of modes supported by the multimode port, the conversion is theoretically lossless. Degrees of freedom are therefore conserved and the second law of thermodynamics (brightness theorem) is not violated~\citep{mcmahon75}. A copy of the IO-based astrophotonic device, which tends to be inexpensive to replicate, is then needed at the output of every SMF in order to recover all the collected flux. The number of modes required to efficiently couple starlight at a telescope's focus scales as the square of the aperture diameter.  
This results in $\sim100$s of modes being required and consequently $\sim100$s of single-mode channels at the output of the photonic lantern (almost $1000$ modes for a $4$ m telescope at median seeing). While such complex lanterns can, in theory, be fabricated, the total flux divided among too many channels will result in every SMF having a fractional share of the total optical power comparable to, or even less than, the noise floor of the detector. Accumulating these noisy signals in post-processing would result in a signal-to-noise ratio (SNR) smaller than that had all the flux been collected by a sole SMF directly from the focal plane to the instrument.     

\begin{figure}
	\includegraphics[width=\columnwidth]{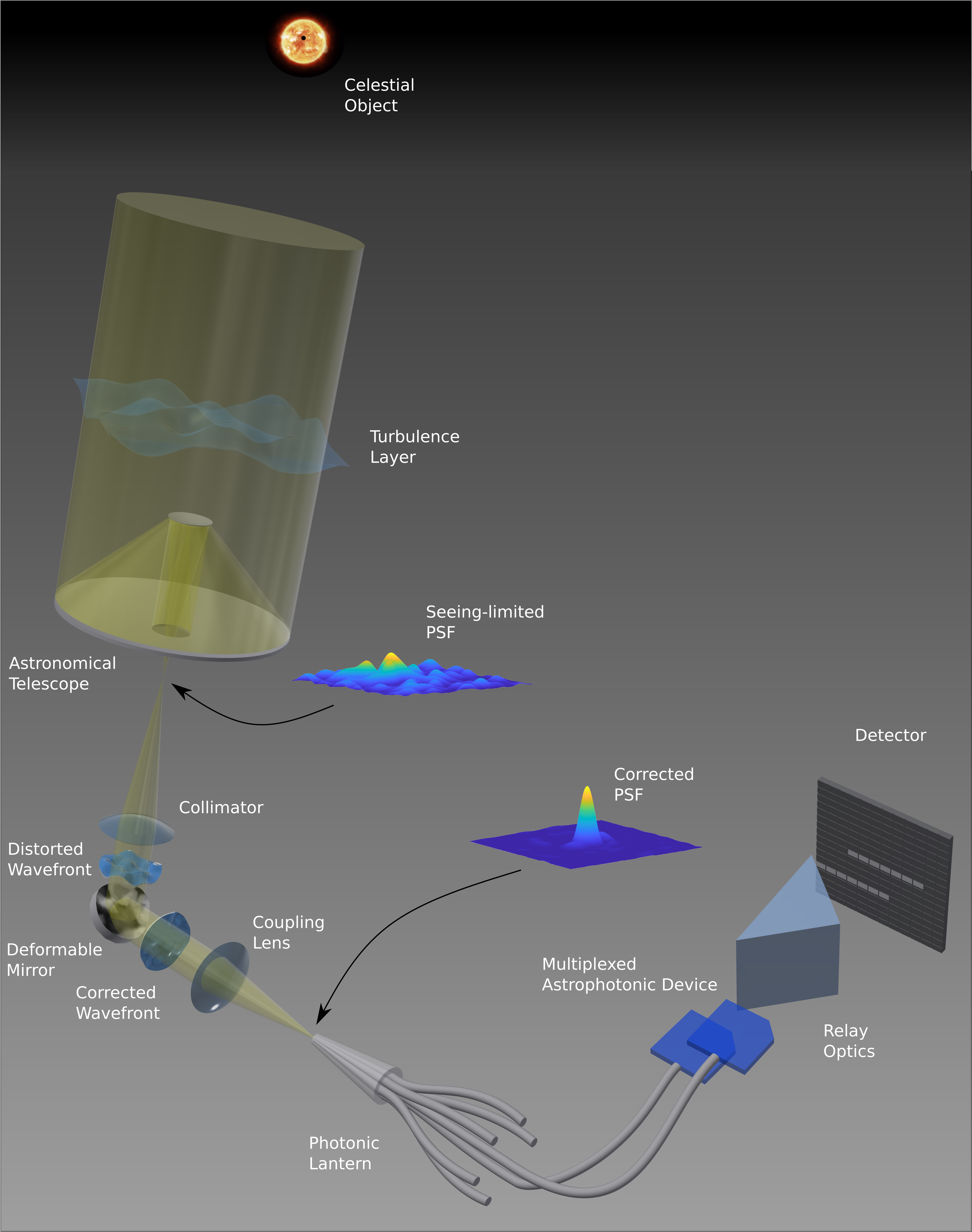}
    \caption{Concept overview. Starlight collected by an astronomical telescope is corrected by an AO system before getting coupled into a photonic lantern. Replicas of an astrophotonic device at the SMFs tips manipulate copies of the same signal to generate multiple images at the detector that can be stacked in post-processing.}
    \label{fig:concept}
\end{figure}

An alternative approach is to combine the two techniques, i.e. AO and photonic lantern,  as illustrated in Fig.~\ref{fig:concept}. Here, the goal is to partially correct the incident wavefront using a low-order adaptive optics (LOAO) system first to reduce the modal content down to a manageable number ($\sim 10$s) before coupling the starlight into the multimode port of a reasonably-sized photonic lantern where a multiplexed photonic spectrograph like the instrument suggested by~\citet{watson1995} and PIMMS~\citep{pimms} can be used at the single-mode output ports. The same signal, e.g. spectrum, is thus measured multiple times. Such a hybrid solution is potentially less expensive than employing an ExAO system and amplifies the SNR compared to a standalone photonic lantern. To find the optimum trade-off between the complexity, i.e. degrees of freedom, of the LOAO system, and the number of modes, i.e. degrees of freedom, of the photonic lantern, a study of how an LOAO system affects the form of the PSF for various turbulence strengths and how coupling efficiency into fibers 
depends on the number of modes, is needed. 
The trade-off can vary for different instruments: depending on readout noise (RON) and other detector properties, an optimum number of SMFs exists for a given LOAO system such that the SNR of the accumulated signal is maximized.

Coupling through turbulence directly into SMFs  has been studied for both astronomy and free-space optical (FSO) communication applications~\citep{ruilier, shaklan, dikmelik}.~\citet{horton} calculated coupling into FMFs numerically for the diffraction-limited case while~\citet{zheng} explored coupling via seeing-limited telescopes but only up to $4$ modes. Coupling into a $1\times7$ photonic lantern of high Strehl ratio PSFs was demonstrated on-sky using the ExAO available at the Subaru telescope~\citep{Subaru} and experimentally without correction for FSO scenarios by~\citet{tedder20}.

Here we study systematically for the purposes of H-band astronomy the dependence of throughput and SNR on the turbulence strength, the extent of AO compensation, the number of modes sustained by the coupling wave\-guide, the setup geometry (its f-number) and the detector quality. Section~\ref{sec:methods} revisits the basic physics and the mathematical tools used to obtain the necessary models for the atmospheric layer, the AO system, and the wave\-guides considered. The simulations run utilizing those models to calculate the dependency of coupling efficiency on f-number and turbulence strength are described in Section~\ref{sec:simulations}. An experiment is devised around an LOAO setup to validate the simulation results and check for deviations in the models. Both the experimental setup and the results obtained are presented in Section~\ref{sec:experiments}. 



\section{Methodology}
\label{sec:methods}


The mathematical models used to calculate the coupling efficiency of atmospherically-distorted starlight into fibers are discussed here.
First, the wavefronts are propagated through a ground atmospheric layer before getting clipped by an entrance pupil. The corrugated wavefronts are then passed through an AO system to calculate the partially corrected beams and their residual error. Next, the corrected wavefronts are propagated to the focus where the coupling efficiency of the collected light into SMFs and FMFs is calculated using the supported modes of the wave\-guides.

\subsection{Atmospheric turbulence}
\label{atmosphere}

Distant celestial objects observed through apertures appear as point sources emitting electromagnetic waves of planar phase fronts and uniform intensities at the top of Earth's atmosphere. Upon propagating through the turbulent atmosphere, the optical field's phase and amplitude are distorted before reaching ground-based telescopes. The distortion in phase~\citep{roddier_adaptive_2008} results in the deformation of the PSF from the diffraction-limited Airy pattern, where $84$ per cent of the total power is within one central disk, into a speckle pattern where the collected power is spread among many loci, the number of which depends on the state of the atmosphere and the diameter of the collecting telescope (see Fig.~\ref{fig:concept}). The weak overlap of such a speckle pattern with the first few modes of a step-index fiber means that coupling starlight efficiently into a narrow wave\-guide cannot be achieved without compensating for the atmospheric turbulence or increasing the number of modes that the fiber supports. An LOAO system can partially correct the corrugated wavefront prior to coupling because it has the effect of redistributing most of the optical power from the speckles back into a central core, albeit with a background halo, and thereby improving the coupling efficiency into FMFs.

To include the effects of turbulence on light propagating through the atmosphere and subsequent wavefront correction, mathematical models have to be identified and implemented in wavefront calculations.
Realizations of the atmospheric-induced wavefront phase distortion, called phase screens, that have ensemble statistics matching those predicted by Kolmogorov's 1941 theory of turbulence~\citep{kolmogorov1991} are computed~\citep{byron_m._welsh_fourier_1997}. This is a  modal-based representation in which the wavefront is assumed to be a superposition over the aperture of infinite orthogonal basis functions or modes. The basis functions are assigned zero mean Gaussian pseudorandom coefficients that possess the desired variance. In doing so, the modal-based approach avoids the shortcomings of sample-based methods, namely the underrepresentation of low spatial frequencies.

An LOAO system is a single conjugate adaptive optics (SCAO) system of modest capabilities. SCAOs are limited to corrections of a single atmospheric layer, which in this case is taken to be the ground layer at the pupil where most of the distortions occur.  
The phase screens simulating the ground turbulence layer are non stationary random functions that can be described by structure functions introduced by Kolmogorov 

\begin{equation}
    D_{\phi}(\rho) = \langle\lvert\phi(\vec{r}) - \phi(\vec{r}+\vec{\rho})\rvert^2\rangle,
	\label{eq:SF}
\end{equation}

where $\phi(\vec{r})$ is the phase at a point located by the vector $\vec{r}$ and $\phi(\vec{r}+\vec{\rho})$ is the phase at a point a distance $\rho = \lvert \vec{\rho} \rvert$ away. The von K\'arm\'an power spectral density (PSD) associated with this structure function is~\citep{hardy_adaptive_1998}

\begin{equation}
    \Phi(\kappa) = 0.023 \left| \kappa^2 + \frac{1}{\mathcal{L}_0^2} \right|^{-11/6}r_0^{-5/3},
	\label{eq:psd}
\end{equation}

where $\kappa$ is the spatial frequency, $\mathcal{L}_0$ is the outer scale, and $r_0$ is the convenient Fried parameter~\citep{fried65} that quantifies the accumulated turbulence strength over the thickness of the turbulence layer, defined as

\begin{equation}
    r_0 = \left[0.423 k^2 \sec{\gamma} \int_0^{\infty} C_n^2(h)dh\right]^{-3/5},
	\label{eq:r0}
\end{equation}

where $k$ is the wavenumber, $\gamma$ is the zenith angle and $C_n^2(h)$ is the refractive index structure constant at height $h$ above the aperture. The total wavefront variance $\sigma_{\phi}^2$ in terms of turbulence strength is $1.03 (D/r_0)^{5/3}$. Therefore $r_0$ defines the telescope aperture of diameter $D$ over which the variance in phase $\sigma_{\phi}^2 \approx 1$ rad$^2$. The PSD of the von K\'arm\'an spectrum in Eq.~\eqref{eq:psd} differs from Kolmogorov's in one aspect: it assumes a finite outer scale $\mathcal{L}_0$ and thus suppresses the contribution of frequencies lower than $1/\mathcal{L}_0$. In doing so, it avoids the infinite power in Kolmogorov turbulence as $\kappa \rightarrow 0$. Values for $\mathcal{L}_0$ differ for different observation sites and measurements in the literature disagree widely, but in general it has a value in the range $1 \sim 100$ m, meaning an exact value is superfluous for small telescopes as it only amounts to an overall tilt~\citep{hardy_adaptive_1998}. In this work we assume $\mathcal{L}_0 = 20$ m, an average of the values measured for the Mauna Kea Observatory ($20$ m), the Observatoire de Haute Provence ($23$ m)~\citep{maire07}, and the Palomar Observatory sites ($17.5$ m)~\citep{ziad04}. Due to the random nature of atmospheric turbulence, all calculations affected by it need to be taken as averages over a large number of screens. Here, $85$ screens are used which is found to produce a relative standard deviation in the metrics in question that is lower than unity. The fractal nature of Kolmogorov's phase screens allow for scaling the aperture size down to convenient diameters since the statistics remain the same inside the inertial range $[\ell_0, \mathcal{L}_0]$, apart from a scale factor~\citep{lane1992}.   

\subsection{Partial adaptive optics correction}
\label{sec:ao}

To include the effect of partial AO correction on the corrugated wavefronts, the influence functions of an ALPAO DM97-15 are used to model the deformable mirror (DM). The DM is highly linear ~\citep{gorkom2018}, i.e. its actuators are decoupled and the influence functions completely characterize its behavior. With $97$ actuators arranged inside an octagon, $11$ actuators are on the longest axis across the diameter. Projected on a midsize $4$ m telescope, each DM segment would have a projected size $d = 0.36$ m on the entrance pupil. Since a DM only corrects for optical path delays between its actuators and cannot correct those within an actuator's action area, this configuration 
can only attain the diffraction limit, $\sigma_{\phi} < 1$ rad, for seeing conditions $r_0 > 0.36$ m.  At median seeing conditions for H-band astronomy ($r_0 \sim 0.1$ m), the mean square fitting error is $\sigma^2 \approx 3$ rad$^2$ (for a continuous facesheet DM)~\citep{miller03}; ergo, the correction is partial~\citep{hardy_adaptive_1998}. Qualitatively, the PSF of such a partially corrected wavefront has a central core with an angular radius $\sim \lambda/D$ and a background halo with  an angular radius $\sim \lambda/r_0$. 

\begin{figure}
        \centering
        \includegraphics[width=\columnwidth]{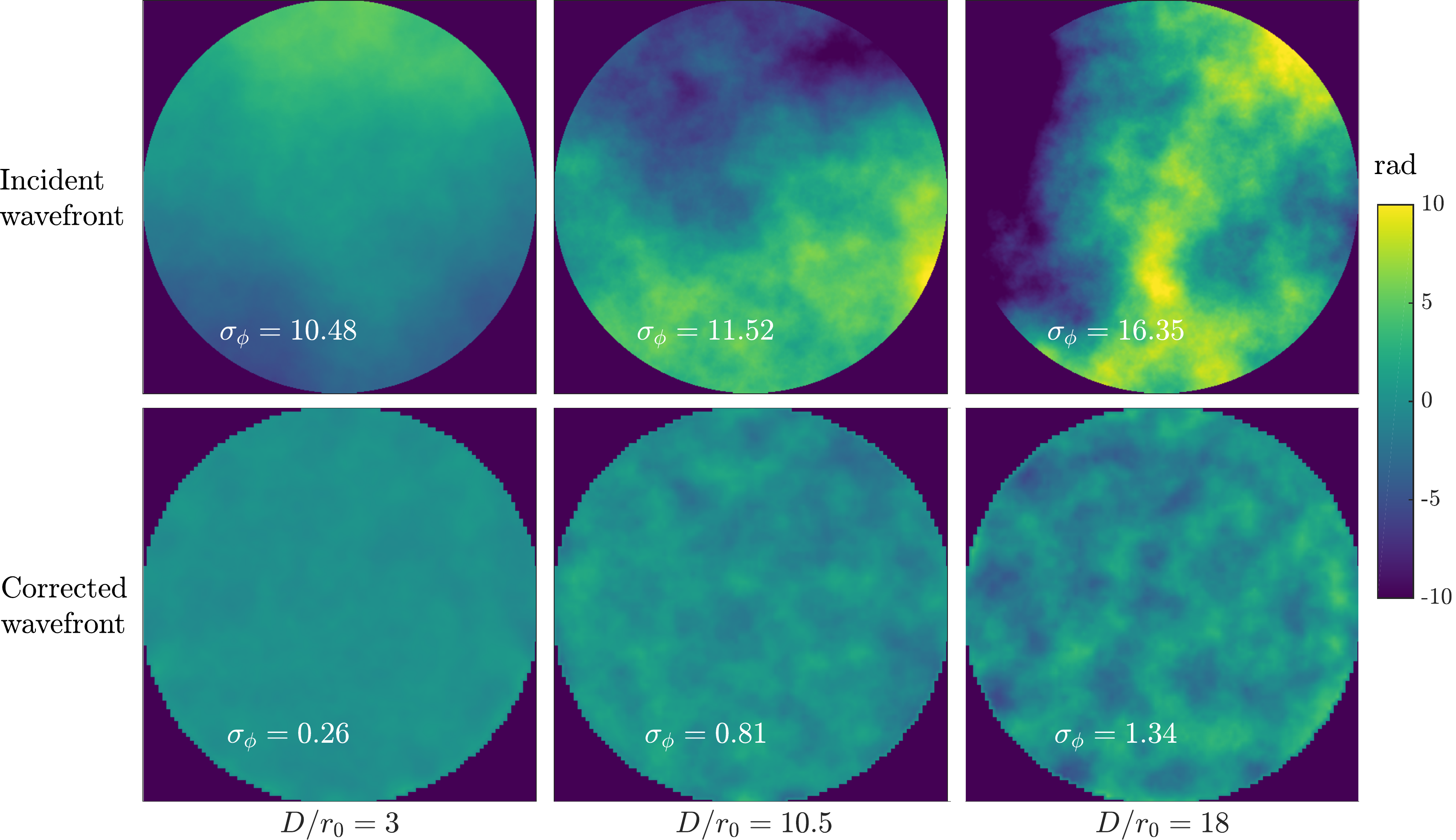}

    \caption{Top: piston-compensated phase screens that exhibit Kolmogorov statistics at varying turbulence strengths. Bottom: corresponding corrected wavefronts by a $97$ degrees of freedom LOAO system.}
    \label{fig:WFs}
\end{figure}

To calculate the commands for the DM, the wavefront is propagated through a model of a Shack-Hartmann wavefront sensor (SH-WFS) that has $10\times10$ subapertures. With actuators at the corners of subapertures, this represents a Fried geometry. Instead of phase, local wavefront slopes inside the subapertures are sensed by the SH-WFS. 
A modal reconstruction is therefore necessary and is performed with the DM influence functions as a basis to calculate the actuators commands

\begin{equation}
    \vec{c} = \mathbf{B}^{+} \vec{s},
	\label{eq:AO}
\end{equation}

where $\vec{c}$ is a vector that contains the DM commands and $\mathbf{B}^{+}$ is the Moore-Penrose pseudoinverse of the matrix $\mathbf{B}$ that has the WFS response to the influence functions. The system in Eq.~\eqref{eq:AO} is underdetermined and therefore lacks a unique solution. The pseudoinverse calculates the solution with the least square departure from a linear fit by performing a singular value decomposition (SVD) that sets the wavefront modes that have a small WFS response to zero. The vector $\vec{s}$ contains the $x$ and $y$ slopes of the incident wavefront calculated from the center-of-gravity of the focal spots behind the microlens array (MLA). Taking advantage of the DM linearity, the shape of the DM due to the commands $\vec{c}$ can be calculated by

\begin{equation}
    \vec{M} = \mathbf{I}^{\intercal} \vec{c},
	\label{eq:AO2}
\end{equation}

where $\mathbf{I}$ is the matrix containing the influence functions of the DM and $^\intercal$ denotes a transpose.  Example realizations of wavefronts generated as described in Sec. \ref{atmosphere} and their corrected counterparts are shown in Fig. \ref{fig:WFs}.

To calculate the coupling into fibers and DM commands, the conjugate counterparts to the entrance pupil (aperture function) at the telescope focus and at the MLA focal plane are required.
Propagating optical fields from pupil planes to focal planes (far-field) is performed by means of the Fraunhofer diffraction equation. A fast Fourier transform (FFT) of the pupil field produces the focal field apart from a coordinate transformation and a scaling phase pre-factor that needs to be included. The optical field at the focal plane, a distance $f$ away, is given by ~\citep{goodman2004}

\begin{equation}
    \psi_F(x,y) = \frac{1}{i\lambda f}\exp\left[i\frac{2\pi}{\lambda}\left(f+\frac{x^2+y^2}{2f}\right)\right] \mathfrak{F}\{\psi_P(x',y')\}_{k_x,k_y},
	\label{eq:Frauenhofer}
\end{equation}

where $\mathfrak{F}\{\psi_P(x',y')\}_{k_x,k_y}$ is the Fourier transform of the pupil field $\psi_P(x',y')$ evaluated at $k_x = x/(\lambda f)$ and $k_y = y/(\lambda f)$. To perform a back-propagation from the focal plane to the pupil plane, an inverse Fourier transform is calculated with the reciprocals of the phase pre-factor and the spatial frequencies instead.  

\subsection{Waveguides}

The models for the SMFs and FMFs 
assume weakly guiding wave\-guides, i.e. low index contrast between the core ($n_{co}$) and the cladding ($n_{cl}$) such that $\Delta n = n_{co} - n_{cl} \ll n_{cl}$,  where both the electric and magnetic fields are transverse to the optical axis. 
Furthermore, circularly symmetric, step-index wave\-guides are assumed which allow the approximation of the guided waves by the linearly polarized (LP) modes. Given the wavelength, the refractive indices, and the core diameter of the fiber, the LP modes are analytically calculable~\citep{saleh} and represent a complete model for the straight fiber. The number of modes that the fiber supports depends on the normalized frequency $V = 2\pi a \mathit{NA}/\lambda$ where $a$ is the core radius and $\mathit{NA}$ is the numerical aperture, $\mathit{NA} = (n_{co}^2 - n_{cl}^2)^{1/2}$. An SMF has $V < 2.405$ and the number of modes $p$ in a multimode fiber (MMF) scales with $V$ according to $p \approx V^2/4$ for each polarization direction. While this approximation is better suited for larger fibers as FMFs adhere less to geometric optics, it can still be used to estimate 
the parameters of the fiber before solving for the exact modes.

Coupling of starlight collected by a telescope into fibers is calculated by evaluating the overlap integral between the fiber modes and the PSF. For an incident optical field $\psi_E(x,y)$ and the $i-$th fiber mode $\psi_i(x,y)$, the coupling efficiency is

\begin{equation}
    \eta_i = \frac{\lvert \iint dx dy \psi_i(x,y) \psi_E^*(x,y)  \rvert^2}{\iint dx dy \lvert \psi_i(x,y) \rvert^2 \iint dx dy \lvert \psi_E(x,y) \rvert^2} = \frac{\lvert\braket{\psi_i}{\psi_E}\rvert^2}{\braket{\psi_i}{\psi_i}\braket{\psi_E}{\psi_E}},
	\label{eq:etai}
\end{equation}

where the integration window is the smallest area over which neither of the two fields vanishes and the normalization factors in the denominator are to compensate for the possibly unequal total powers contained in the fields. 

Further losses, e.g. bending, insertion, and propagation losses, are accounted for in the experimental results but are not included in the models.   

The number of modes $p$ required to couple a seeing-limited PSF can be derived from the conservation of \'{e}tendue~\citep{minardi_astrophotonics_2020}. In terms of turbulence strength

\begin{equation}
    p \approx \left(\frac{\pi D}{2 r_0}\right)^2,
	\label{eq:N}
\end{equation}

a relation that is more accurate for highly multimode fibers having been derived from geometrical optics considerations. Notice that Eq.~\eqref{eq:N} is nothing but the area of the aperture in units of area elements $r_0\times r_0$.

Besides SMFs and MMFs (including FMFs) we also consider photonic lanterns in our models. Here, we assume photonic lanterns with weakly-guiding
circular step index profiles and the mathematical model described above applies to their multimode ports.  
The propagation of the field from the multimode port to the SMFs through the transition region depends on the modal content of the coupled field and the transversal geometry of the transition region. Beam propagation methods (BPMs) could be used to simulate the photonic lantern and calculate the distribution of the optical power among the SMFs. However the discrete step-by-step calculations involved in BPMs tend to be slow and it is hence unrealistic to run them for a large number of phase screens. Instead, the assumption that the optical power distributes equally among the SMFs is made to model the downstream segments of the photonic lantern. The SNR calculations given below (Sec.~\ref{sec:snr}) are a good approximation as long as all SMFs are receiving comparable shares of the total power. Modal noise and scrambling are discussed further in Sec.~\ref{sec:scramble}

A number of different procedures have been considered to realize the necessary taper transition between the MMF and SMFs necessary for photonic lanterns~\citep{birks_PL}. Inserting a bundle of stripped SMFs inside a capillary, whose index is lower than the refractive index of the cladding, and tapering the stack down using a glass processor (Davenport (in preparation)) is the method of choice in astronomy because throughput is not compromised. In contrast to lanterns made from multicore fibers (MCFs) or using ultrafast laser inscription (ULI) techniques, this method results in a single-mode section with free fibers that can be readily spliced to other components. This however requires the SMFs in the bundle to be arranged in a close pack to maintain symmetry along the taper. With an SMF at the center, $1$, $2$, and $3$ rings of SMFs arranged in a hexagonal lattice, i.e. a centered hexagonal number, result in bundles that have $q = 7$, $19$, and $37$ SMFs, respectively (Davenport (in preparation)). The number of modes sustained by an FMF, $p$, can be controlled by tuning the normalized frequency $V$ where the modes count increments by $1$ as $V$ increases gradually, except for degenerate modes ($LP_{lm}, l \neq 0$) where $2$ modes appear together. Opting for theoretically lossless lanterns with more SMFs than modes at the MM section ($p<q$), two of the FMFs considered for the simulations below have $p = 6$ and $36$ modes, one mode short of the ideal design.       

The LOAO system considered below assumes that tip/tilt are already corrected for by a fast steering mirror (FSM) and that the fiber is aligned at the focus. Aligning SMFs to images of celestial objects at the focal planes of very large telescopes has been achieved with the aid of guiding cameras and can nowadays be done with relative ease~\citep{bechter15}.

\section{Simulations}
\label{sec:simulations}


A calculations pipeline that propagates the wavefronts perturbed by the atmosphere from the telescope pupil, through the LOAO system to the focus where a fiber is placed, is built using the mathematical tools in Sec.~\ref{sec:methods}. Estimates for the optimum setup geometry to couple into SMFs and FMFs under diffraction- or seeing-limited conditions can thus be calculated. The boost in coupling that an LOAO system provides as compared to the uncorrected case is studied for various turbulence strengths. The optimum number of channels in a multiplexed astrophotonic device fed by a photonic lantern can therefore be deduced from this pipeline. 

In the following subsections, several free parameters are varied to study the dependencies of the coupling efficiency and the SNR before an optimum number of channels is selected for specific cases. The free parameters under study are the telescope properties in form of the f-number (Sec.~\ref{sec:f-number}) as well as turbulence properties (Sec.~\ref{sec:turbu}). The theoretical SNR of the stacked signal detected at the outputs of a photonic lantern is eventually
calculated to estimate the optimum trade-off in the size of a multi-channel astrophotonic instrument (Sec.~\ref{sec:snr}). Additionally, scrambling is discussed in further detail in Sec.~\ref{sec:scramble}.

\subsection{Coupling dependence on f-number} 
\label{sec:f-number}
Coupling of starlight into wave\-guides depends on the correlation between the PSF and the modes of the wave\-guide. The PSF is defined by the shape of the aperture and the optical field at the pupil. The linear extent of that pattern however is dependent on the effective focal length $f$. As a result, the coupling efficiency depends on the f-number \textsf{\textflorin}$/\# = f/D$ of the coupling setup. For a large MMF, one simply considers the ray optics aspect and selects an \textsf{\textflorin}$/\#$ that gives a beam slow enough to fit within the acceptance angle, i.e. $\mathit{NA}$, 
of the fiber. Diffraction effects are still non-negligible for SMFs and FMFs, and calculating the coupling efficiency as a function of the \textsf{\textflorin}$/\#$ around the value predicted by geometric optics is necessary to optimize the system.

\begin{figure}
        \centering
        \includegraphics[width=\columnwidth]{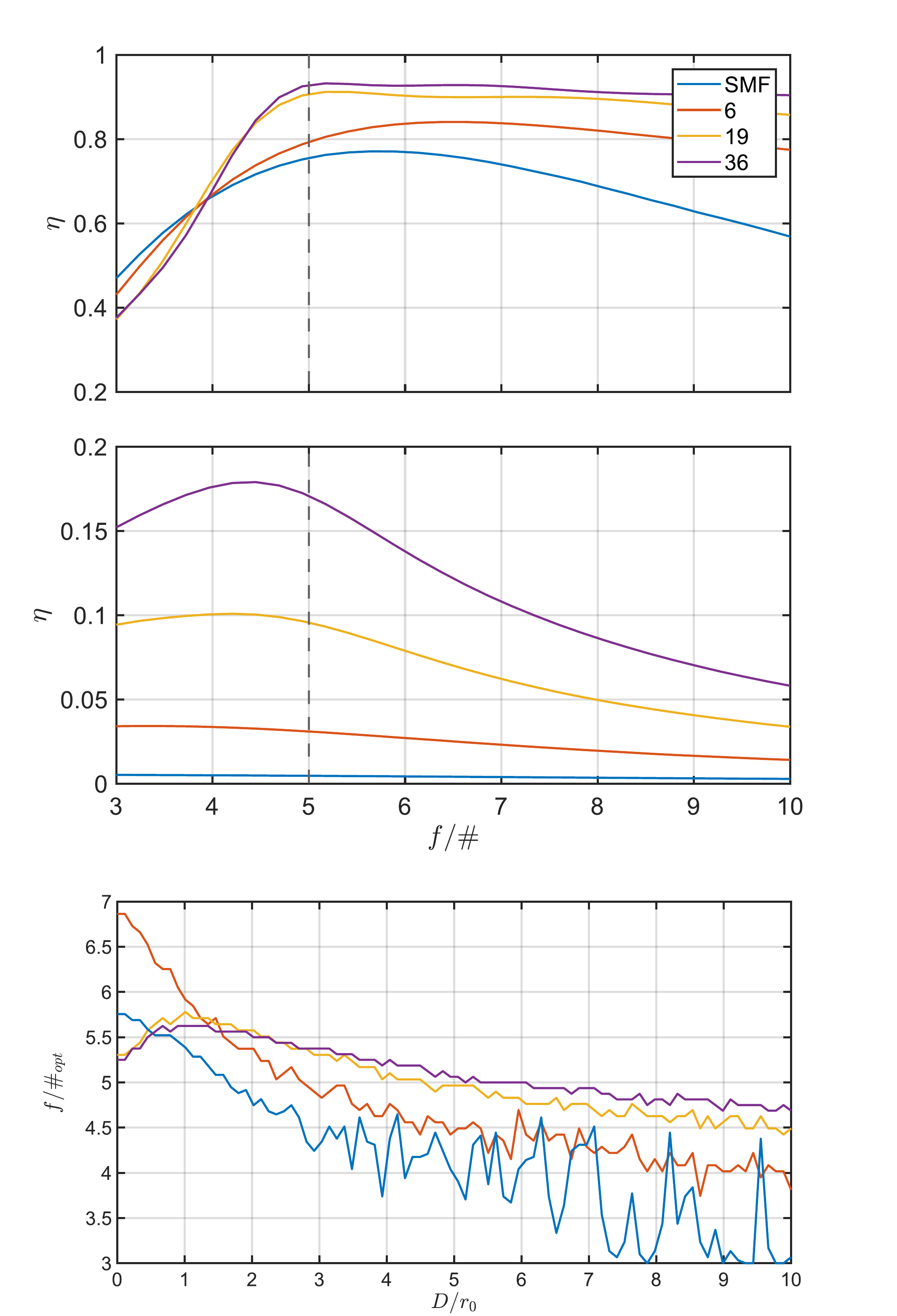}
        \label{fig:eta_fno_DL1}
    \caption{Top: coupling efficiency dependence on \textsf{\textflorin}$/\#$ for $\mathit{NA} = 0.1$ SMF, $6$, $19$, and $36$ modes FMFs at the diffraction limit. Dashed line indicates the geometrically predicted optimum \textsf{\textflorin}$/\#$. Middle: shows the same efficiency dependence, but with added $D/r_0 = 15$ turbulence. Bottom: variation in optimum f-number as turbulence increases.}
    \label{fig:eta_fno_DL}
\end{figure}



Fig.~\ref{fig:eta_fno_DL} shows the dependence of the coupling efficiency $\eta$ on the f-number \textsf{\textflorin}$/\#$ for the SMF and FMFs sustaining $p = 6$, $19$, and $36$ modes at $\lambda = 1550$ nm. All fibers have $\mathit{NA} = 0.1$ from which the geometrically expected optimum value is \textsf{\textflorin}$/\# \approx (2\mathit{NA})^{-1} = 5$. 
The efficiency increases rapidly and reaches its maximum values at \textsf{\textflorin}$/\#$ around the geometrically predicted values (\textsf{\textflorin}$/5.1$ for $p = 36$). For higher $p$, the efficiency curves plateau for slower beams, but with an additional oscillation. The maximum efficiency values increase with increasing $p$, from $0.78$ for the SMF up to $0.91$ and $0.92$ for $p = 19$ and $36$ modes, respectively.
 Fig.~\ref{fig:eta_fno_DL} (middle) shows the expected lower $\eta$ for a seeing-limited case $D/r_{0} = 15$ without AO correction. $\eta$ increases gradually , and following a maximum, gradually decreases instead of plateauing. 
The position of the maxima for $p = 19$ and $36$ mode FMFs, however, are very close to those in the diffraction-limited case (for $p = 36$ modes, the maximum efficiency is reached at \textsf{\textflorin}$/5.1$ and \textsf{\textflorin}$/4.9$ for the diffraction- and the seeing-limited case, respectively). The dependence of \textsf{\textflorin}$/\#_{\mathit{opt}}$ on $D/r_0$ for the fibers considered is also shown in Fig.~\ref{fig:eta_fno_DL}. The optimum coupling \textsf{\textflorin}$/\#$ drops as $D/r_0$ increases due to the focal pattern spreading over a larger area and therefore requiring a faster beam to confine its linear extent.  This gradual drop in optimum \textsf{\textflorin}$/\#$ and similarity between the curves suggests that a setup
designed for the SMF and the diffraction limit may be used for the seeing-limited case with only little effect on coupling. The optimum \textsf{\textflorin}$/\#$ established here is used for all subsequent calculations of starlight coupling into fibers.

\subsection{Coupling dependence on turbulence}
\label{sec:turbu}

\begin{figure}
	\includegraphics[width=\columnwidth]{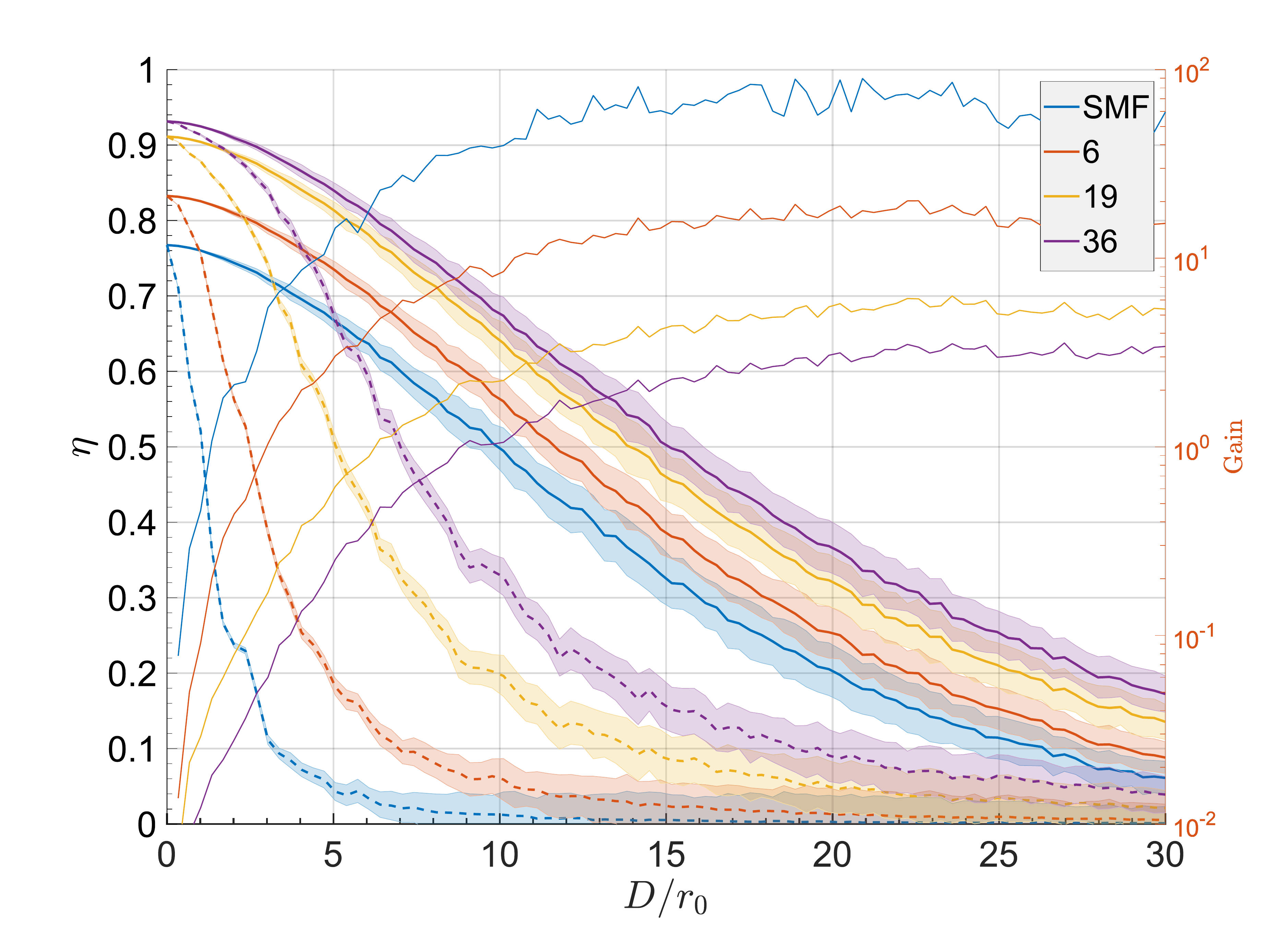}
    \caption{Coupling efficiency dependence on $D/r_0$ for $\mathit{NA} = 0.1$ SMF, $6$, $19$, and $36$ modes FMFs. Solid line: for AO-corrected wavefronts, dashed-line: for uncorrected wavefronts with the shaded bands $2\sigma$ wide centered on the average value. Light solid line: gain in coupling with the log ordinate on the right.}
    \label{fig:eta_dr0}
\end{figure}

The deterioration of coupling efficiency into an SMF and $3$ FMFs as seeing worsens is shown in Fig.~\ref{fig:eta_dr0}. An ensemble of $85$ phase screens was used to calculate the overlap integral at each turbulence strength point between the diffraction limit and $D/r_0 = 30$. The improved coupling efficiency that the partial AO-compensation contributes is also plotted. Phase screens generated with the statistics discussed in Sec.~\ref{atmosphere} are corrected by the LOAO system described in Sec.~\ref{sec:ao} before coupling into the fibers is computed by Eq.~\eqref{eq:etai}. For a $4$ m telescope at $r_{0} = 0.2$ m, $D/r_{0} = 20$ and the level of correction that the LOAO system attains is only partial. The coupling efficiency into the SMF is increased a $100$ fold to about $20$ per cent as depicted in Fig.~\ref{fig:eta_dr0}. A comparable boost in $\eta$ between $20$ and $3$ is attained for the $6$ and $36$ FMFs at $D/r_{0} = 20$, respectively. 
The operation regime for an LOAO assisted SMF- or FMF-fed astrophotonic instrument can therefore be determined from curves like those in Fig.~\ref{fig:eta_dr0}.  

\subsection{SNR dependence on turbulence and modes count}
\label{sec:snr}

In the absence of detector noise as a consideration, maximum throughput could be obtained by using a multi-channel astrophotonic instrument with the highest available number of channels $q$, in order to maximize flux collection. 
Detectors are however always noisy. Splitting the total flux into small shares by a photonic lantern and detecting them separately by a noisy detector before accumulating all the signals in post-processing would only result in a better SNR than using an SMF if the number of channels in the instrument is optimized to the photon flux and the detector's RON. Consequently, the SNR is taken here as a figure of merit to decide on the optimal size of a photonic lantern for a given telescope aperture, seeing condition, target magnitude, AO degrees of freedom, and detector's noise. Sources of noise relevant here are photon shot noise and RON. In a photon-starved application like astronomy, photon shot noise $\sigma_{\mathit{ph}}$ dominates ($\mathit{SNR} = \sqrt{N_{\mathit{ph}}}$, where $N_{\mathit{ph}}$ is the photons count) and delivering more photons via larger FMFs is thus beneficial. 
Moreover, the electronic circuit of the detector used to sense the starlight introduces a constant RON independent of photons count every time a pixel is read out. With the total signal distributed into multiple channels and detected at the outputs of the photonic devices over more pixels, every output signal has an RON component and in the aggregated signal the noise accumulates. The multiplexing approach using a photonic lantern can only yield a better SNR than a standalone SMF-fed device for detectors with minimal RON and bright objects under relatively good seeing conditions.

Astronomy-grade NIR detectors typically have RON values in orders of few electrons 
\citep{finger_saphira_2014} and improved designs for amplifier circuits with sub-electron RON continue to come out~\citep{feautrier16}. The case for a multiplexed H-band astrophotonic device fed by a photonic lantern is therefore driven by the technological advancements of both integrated optics and NIR detectors along with AO.


\begin{figure}
	\includegraphics[width=\columnwidth]{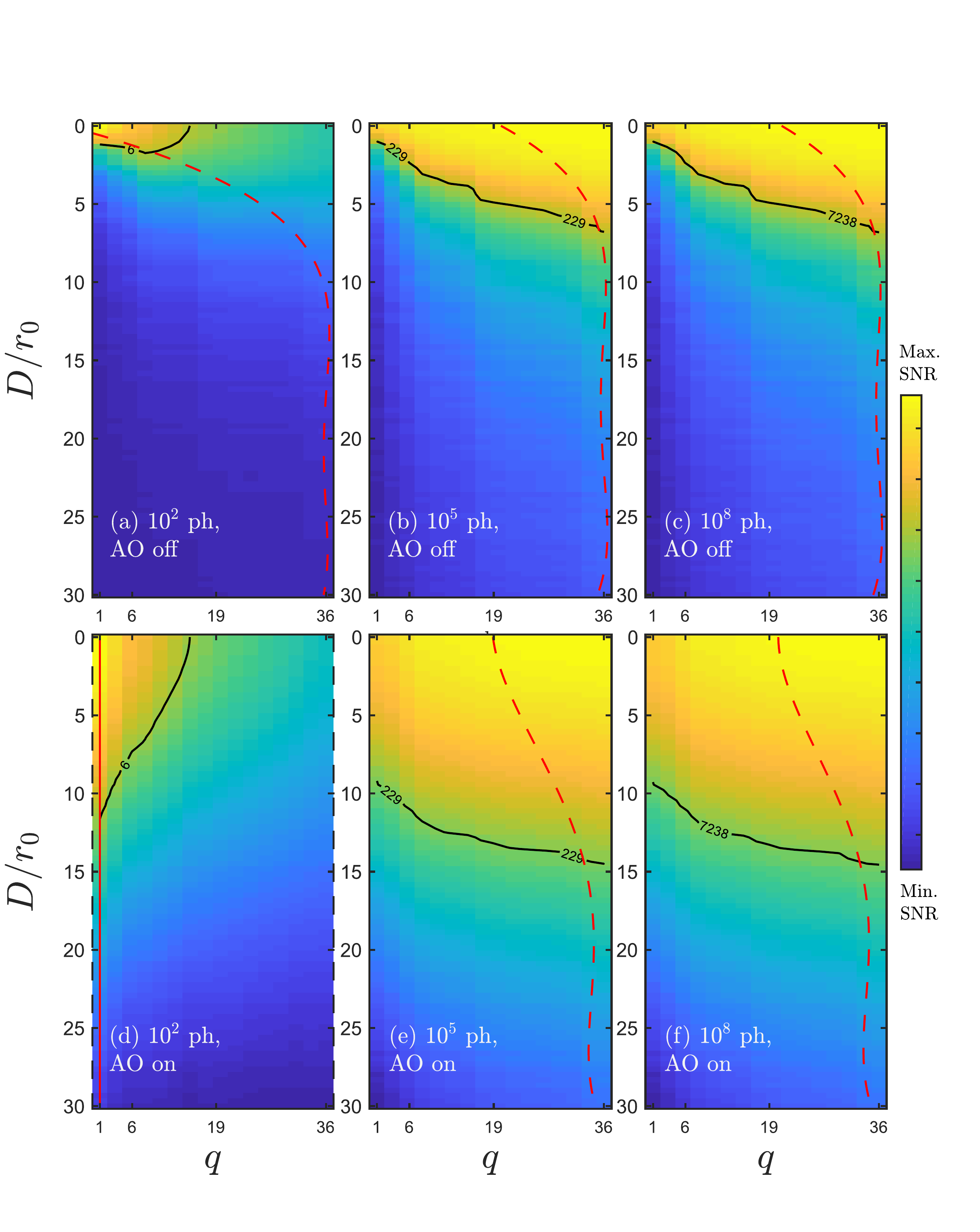}
    \caption{SNR dependence on turbulence strength $D/r_0$ and number of channels $q$ for a detector with RON $\sigma = 3$ e$^-$. Photons count $=10^2$ photons (left panels), $10^5$ photons (middle panels) and $10^8$ photons (right panels). Without AO correction (upper panels) and in the presence of AO correction (lower panels). Black contours are isolines of constant SNR at 0.75 of the maximum achievable for the scenario in question. Dashed red lines are polyfits that trace along the 0.99 of the maximum SNR values for each $D/r_0$ case. }
    \label{fig:SNRs}
\end{figure}

A two parameter calculation of SNR variation with $D/r_{0}$ and the number of detection channels $q$ is performed to decide on the optimum configuration for the multiplexed astrophotonic instrument. With both photon noise and RON considered, SNR for the accumulated signal is

\begin{equation}
    \mathit{SNR} = \frac{\eta \cdot N_{\mathit{ph}}}{\sqrt{\sigma^2_{\mathit{ph}} + q \cdot \sigma^2_{\mathit{RON}}}},
	\label{eq:SNR}
\end{equation}

where $\eta$ is the coupling efficiency and $q$ is the number of channels. Fig.~\ref{fig:SNRs} shows the SNR as a function of $q$ for the example cases of a faint ($N_{\mathit{ph}}=10^2$) and a bright ($N_{\mathit{ph}} = 10^8$) sources detected with a $\sigma_{\mathit{RON}} = 3$ e$^{-}$ detector. For the fainter object, the maximum SNR is attained using a single channel ($q=1$) device for both cases with and without LOAO correction ([a] and [d] in Fig.~\ref{fig:SNRs}). For the brighter objects on the other hand, SNR grows linearly with the number of channels $q$ for all turbulence scenarios in the uncorrected case ([c] in Fig.~\ref{fig:SNRs}) but saturates at around $q = 19$ channels for the LOAO-corrected case ([f] in Fig.~\ref{fig:SNRs}).




\subsection{Scrambling}
\label{sec:scramble}

While all MMFs experience modal noise, FMFs are particularly affected due to the wider separation in terms of effective refractive indices between their supported modes. Scrambling the modes to minimize modal noise might be crucial depending on the application, e.g. high precision radial velocity spectroscopy. For a multiplexed astrophotonic instrument, modal noise entails that different replicas of the IO component at the SMFs of the delivery photonic lantern will receive different amounts of light. 

\begin{figure}
	\includegraphics[width=\columnwidth]{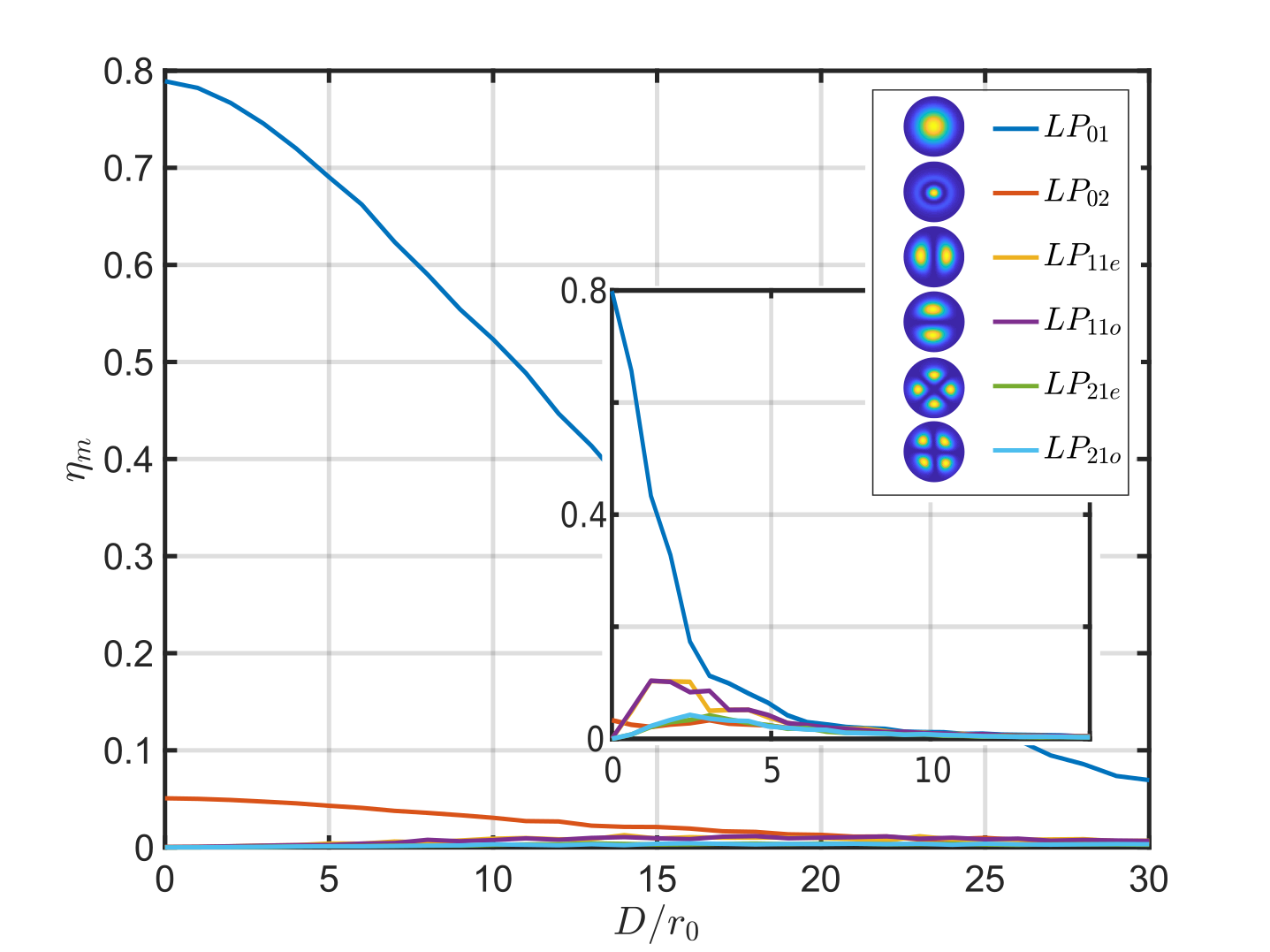}
    \caption{Contribution of individual modes to coupling efficiency into a $6$ modes FMF. Inset shows the uncorrected case. Notice the different ranges on the abscissae. }
    \label{fig:etam_dr0_6_noAO}
\end{figure}

Diffraction-limited and partially AO-corrected PSFs, with most of the optical power in a central disk, excite azimuthally-symmetric ($LP_{0m}$) modes only. As $D/r_0$ increases, more light is present in speckles away from the PSF core and higher, non-circularly symmetric, LP modes are excited with higher probability. 
Fig.~\ref{fig:etam_dr0_6_noAO} shows the coupling efficiency into each mode of a $6$ modes FMF as a function of turbulence strength for the  AO-corrected and the seeing-limited cases. This suggests that shifting the corrected PSF away from the center of the FMF in a controlled manner, akin to the speckles in a seeing-limited focal pattern, can improve scrambling (at the cost of efficiency) as is planned for the NIRPS spectrograph~\citep{wildi2017}. Mechanical agitation~\citep{baudrand01} and stretching~\citep{chen_origin_2006} on the other hand scrambles the light by transferring the optical power that is coupled dominantly in $LP_{0m}$ modes into the other modes along the fiber. Static scrambling strategies, e.g. octagonal fibers, have also been studied~\citep{chazelas2010}.

At the diffraction limit, the IO replica connected to the central SMF (in a hexagonal pack configuration) will get most of the power. Depending on the application, an astrophotonic instrument might be able to deliver a good performance with an unequal distribution of power between the replicas as long as enough light to achieve an $\mathit{SNR} \gg 1$ reaches every replica. Other applications that require an equal splitting of the total flux would require that the light is scrambled among the SMFs using one of the techniques mentioned above. For applications sensitive to modal noise, further investigation of the suitability of the MMFs modelling and scrambling methods for FMFs is needed since the statistical treatment of the fiber noise in large MMFs~\citep{grupp2003} is not appropriate for modelling FMFs.  




The substantial contribution of $LP_{0m}$ modes to coupling as compared to the other modes presents an opportunity for mode-selective photonic lanterns (MSPL). By breaking the symmetry between the SMFs in a photonic lantern, a one-to-one definitive mapping can be enforced between the excited modes and the SMFs~\citep{saval14}. This results in the optical power coupled into the multimode port being guided dominantly to a subset of all the SMFs present, reducing the number of channels and alleviating the effect of the detector's RON. The natural scrambling that takes place due to the atmospheric turbulence limits the utilization of MSPLs for the uncorrected wavefronts case (see Fig.~\ref{fig:etam_dr0_6_noAO}).



\section{Experiments}
\label{sec:experiments}

\begin{figure*}

	\includegraphics[width=1.7\columnwidth]{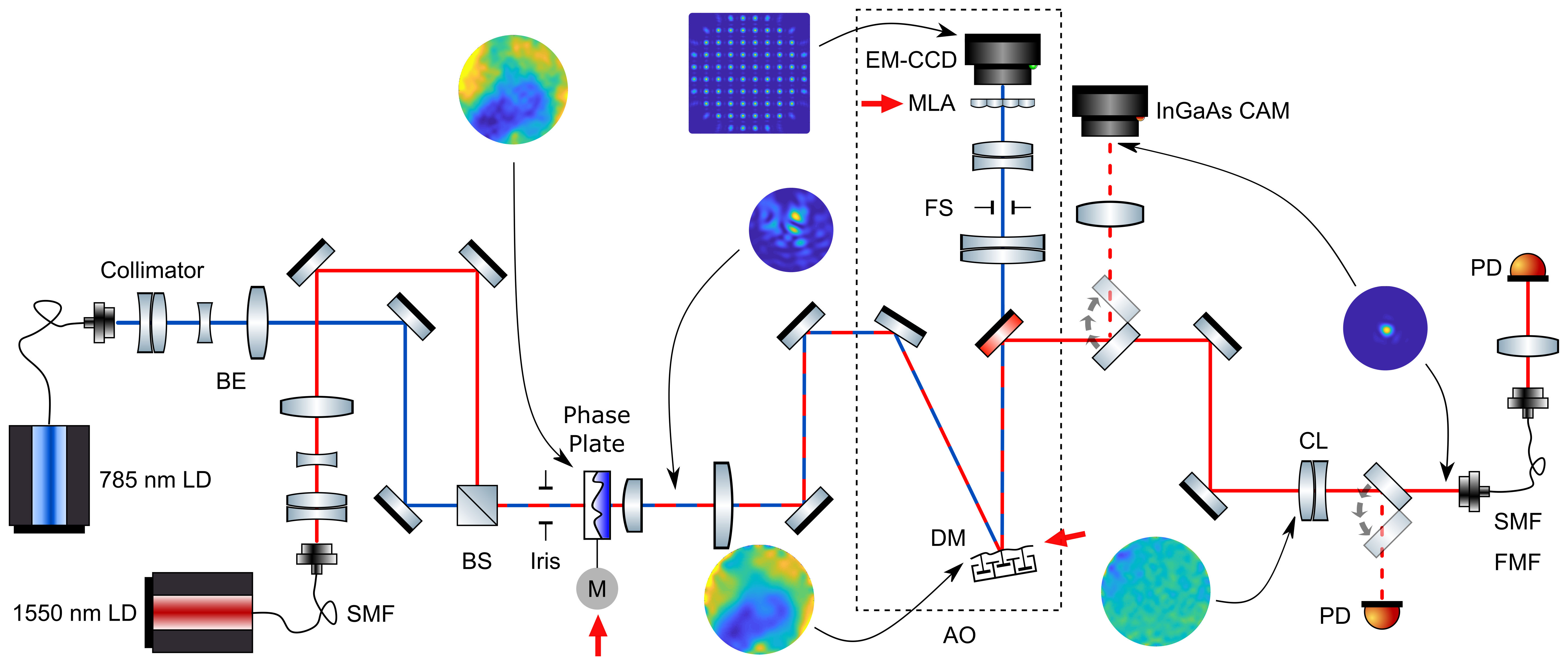}
    \caption{Experimental setup. BE: beam expander, BS: beam splitter, M: stepper motor, CL: coupling lens. Red arrows indicate conjugated pupil planes.}
    \label{fig:experiment}
\end{figure*}

To validate the aforementioned coupling models, the experimental testbed in Fig.~\ref{fig:experiment} is built around an LOAO system.   
The measurements are performed using two light sources, one of which is the science beam at $1550$ nm that simulates star light and a second beam at $785$ nm as guide star for the AO system. The coupling models are tested by recreating the simulated scenarios and measuring the coupling efficiency. The main parts of the experimental testbed are described in the following subsections, including the LOAO system (Sec.~\ref{sec:exp:ao}), the atmospheric emulator (Sec.~\ref{sec:exp:atmo}), the relay and coupling optics (Sec.~\ref{sec:exp:relay}), and the fiber optics (Sec.~\ref{sec:exp:fiber}), with the experimental results presented in Sec.~\ref{sec:exp. results}.

\subsection{Atmosphere emulator - Phase screen}
\label{sec:exp:atmo}


The two beams that emulate the astronomical science target and the guide star are provided by a $1550$ and a $785$ nm fiber-coupled laser diodes, respectively. The beams are collimated by doublet achromats and enlarged by Galilean beam expanders to $24$ mm diameters. Light polarization is not maintained. 
The beams are reflected off folding mirrors that steer them toward a beam splitter where the visible beam is transmitted and the NIR beam is reflected onto a common axis (c.f. Fig.~\ref{fig:experiment}). The combined beams are passed through a phase screen mounted on a rotary stage. The phase screen from LEXITEK is a $100$ mm diameter plate of two polymers of similar but unequal refractive indices bonded together~\citep{ebstein96}. The near index match between the two materials means that the physical profile variation across the screen to produce a given optical path difference is coarser than that for an air-glass interface. This relaxation results in a profile difference as high as $75$ $\mu$m being required to introduce a phase shift of only $1$ wave on an incoming planar wavefront at $1550$ nm. This allows for engraving the phase pattern using typical CNC machining techniques. The sandwich is itself cemented between two $\lambda/10$ BK-7 windows with AR coating. All materials are transparent between $600$ nm and $1600$ nm and the dispersion in the refractive index difference $\Delta n$ of the two polymers is less than $0.001$ between $785$ nm and $1550$ nm. The phase pattern is impressed on an outer annulus of width $24$ mm and thus a maximum beam diameter of $24$ mm can be passed through the plate. The plate is mounted off-center to overlap the outer annulus with the beam path, and different pattern realizations can be attained by rotating the plate about its axis. 
The phase screen engraved has Kolmogorov statistics with an $r_0 = 0.6$ mm. 
The beams emerging from the phase plate are truncated by an iris to allow the emulation of different telescope aperture sizes. With the fixed Fried parameter of the screen, varying the turbulence strength up to a maximum of $D/r_0 = 40 $ is possible by changing the iris opening diameter. 

\subsection{Low-order adaptive optics}
\label{sec:exp:ao}


The LOAO system from ALPAO has a $13.5$~mm DM with $97$ actuators, that can achieve a flatness of $2.89$~nm rms in closed loop. It can be driven up to a  frequency of $1$~kHz before the first resonance sets in. Wavefront sensing is performed downstream of the DM in a closed-loop architecture by a Shack-Hartmann type sensor that consists of a $16 \times 16$ MLA with an electron-multiplying CCD (EM-CCD) camera. The EM-CCD from N\"uv\"u has a quantum efficiency $\mathit{QE} > 75\%$ between $500$ and $800$ nm and 
a full frame rate of $1$ kHz. The LOAO system features a dichroic mirror to split the science beam from the guide star beam which in turn is resized by a beam reducer (a Keplerian telescope of $2$ achromats) to match the size of the lenslet array ($2.75$ mm). The beam reducer also conjugates the DM pupil to the lenslet array that is placed at the exit pupil of the system. A field stop at the image plane between the telescope lenses allows for blocking the light from all sources other than the guide star beam. The loop is controlled with a PC driven by an Intel i5-8500, $3.0$ GHz, $6$ cores processor and $16$ GB RAM running a Matlab-based engine developed by ALPAO. The pure delay, i.e. the time it takes the PC to have the DM commands ready after the end of an exposure, is $1.38$ ms. The loop can be closed at $500$ Hz with the WFS camera operating at maximum frame rate with sufficient photons available for each frame to guarantee a high signal-to-noise ratio (camera requires $3$ photons/frame/subaperture for SNR $=1$). The rejection (correction) bandwidth, i.e. the maximum Greenwood frequency that can be corrected for, is the figure of merit most important for our purposes and for this system it is about $20$ Hz. The partial correction nature of the LOAO system is clear from the limited number of modes ($97$) it can correct for (spatial limitation) and the maximum Greenwood frequency it can track (temporal limitation) compared to the needs of large and very large telescopes.

\subsection{Relay optics and coupling setup}
\label{sec:exp:relay}


After passing the phase screen and the iris, the clipped beams are then passed through a Keplerian telescope that resizes the beam diameter to match the DM aperture. The secondary lens of the telescope also images the phase screen at the primary lens, i.e. the entrance pupil, onto the DM when the telescope is kept at the correct image distance from the DM. The telescope lenses are uncoated singlets, and lens pairs that produce the minimum total Seidel aberrations were selected from off-the-shelf catalogues using Zemax. The exit pupil position is also calculated by Zemax.


The combined beams are then folded towards the DM, where the wavefront is actively controlled. After reflection off the DM, the AO-corrected beams are split by a dichroic beam splitter (c.f. Fig.~\ref{fig:experiment}). While the guide star beam goes to the WFS, the science beam is reflected out of the AO setup toward a coupling lens with an SMF or FMF aligned at its focus. The same $f=75$~mm lens is used for all fibers. With the beam diameter at the lens equal to the DM aperture, i.e. $d = 13.5$~mm, an \textsf{\textflorin}$/5.56$ beam results which is close to the optimum for coupling into an $\mathit{NA} = 0.1$ SMF at $\lambda = 1550$~nm as shown in Fig.~\ref{fig:eta_fno_DL}. The optimum \textsf{\textflorin}$/\#$ for coupling into SMFs of a different $\mathit{NA}$ or FMFs is different, but $5.5$ is roughly the midpoint where sufficient coupling is expected for all the cases considered. 

To measure the total power coupled, the output end of the fiber is connected to a fiber port where the facet is imaged onto a free-space photodetector (PD) by a singlet lens. Such an arrangement is necessary because the femto-watt PD has a small active area ($0.2$ mm$^2$), which makes it difficult to completely capture the diverging beam directly after the fiber.
The highly sensitive InGaAs PD has a noise equivalent power $\mathit{NEP}= 7.5$ fW and bandwidth $\mathit{BW}= 25$ Hz.

The NIR beam is intercepted at two points by flip mirrors. First, before the coupling lens, the beam can be sent toward a C-RED2 InGaAs camera~\citep{feautrier2017} through a long focal length lens $f = 1000$~mm to image the PSF and measure its Strehl ratio and encircled energy. Second, after the coupling lens and before the fiber, the beam can be reflected towards a free-space highly sensitive PD to measure the total power available at the pupil for coupling. The two PDs were calibrated against each other using the fiber port-lens-PD setup across their dynamic range $10$~fW - $100$~pW. 

Flip mirrors are used instead of beamsplitters after separating the guide star beam from the science beam
to reduce the number of surfaces that may give rise to non-common path aberrations (NCPA) between the AO setup and the fiber. To align the fiber at the focus of the coupling lens, a 6-axis kinematic precision stage is used. 
The nanopositioning stage used can achieve a resolution of less than $50$~nm for $x$, $y$, and $z$. Pitch and yaw have a $0.2$~arcsec resolution.   


A feedback loop is closed between the C-RED2 camera and the DM (on top of the main AO loop) to correct for temperature-induced tip and tilt in the IR arm only and hence not seen by the WFS. Wavefront error, temperature and humidity are logged for the open loop case to make sure that creep and memory effects as reported by~\citet{bitenc2014} do not introduce measurement artifacts.

\label{sec:exp:fiber}


Coupling is measured for an SMF and an FMF. 
A coated SMF-28 patch cable from Corning is used as the SMF. For the FMF, a THORLABS FG025LJA MMF is used. Table~\ref{tab:fiber} lists the properties of the fibers 
used.
Short fibers ($<2$ m) are used to minimize attenuation in the fibers and cladding modes are removed by having the fibers bent at radii ($>30$mm) larger than their macrobend loss thresholds. 

\begin{table}
	\centering
	\caption{Properties of the fibers used.}
	\label{tab:fiber}
	\begin{tabular}{lccr} 
		\hline
		Fiber & Core diameter [$\mu$m] & $\mathit{NA}$ & Sustained modes \tablefootnote{Per polarization and accounting for degeneracies at $\lambda = 1550$ nm}\\
		\hline
		SMF\tablefootnote{Fiber is AR-coated at one end} & 8.2 & 0.14\tablefootnote{Measured at $1$ per cent power level} & 1\\
		FMF & $25\pm1$ & $0.08\pm0.005$ & 6\\
		\hline
	\end{tabular}
\end{table}

\subsection{Experimental results}
\label{sec:exp. results}
\begin{figure}

	\includegraphics[width=\columnwidth]{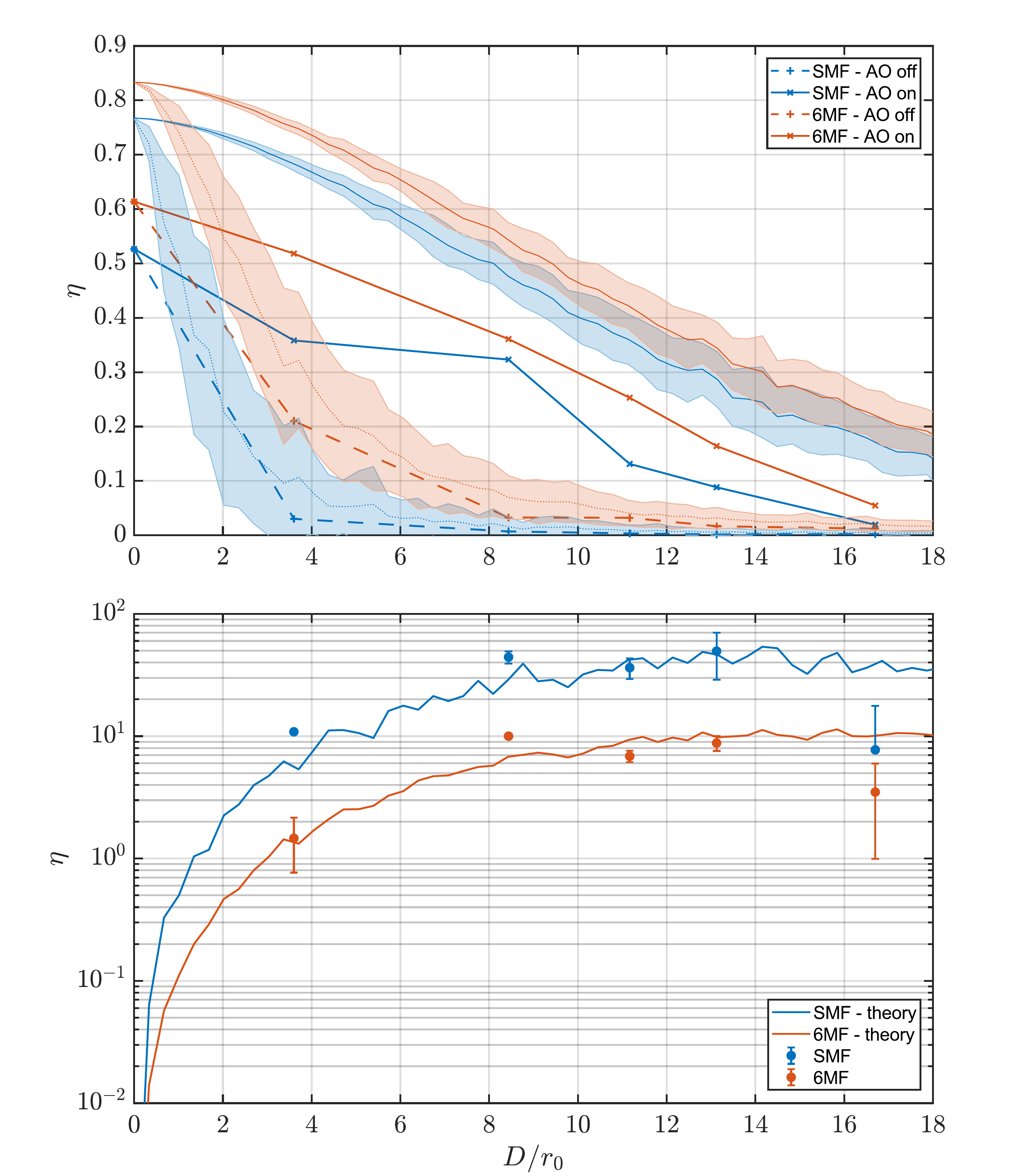}
    \caption{Top: measured coupling efficiency into SMF and $6$ modes FMF. For reference, the simulation results are plotted (light lines with shaded areas indicating uncertainties). Bottom: measured gain in coupling upon application of AO correction. The line plots going through the data are the simulation results.}
    \label{fig:exp. res}
\end{figure}

Fig.~\ref{fig:exp. res} shows the measured coupling efficiencies and gain for an SMF and a $6$-modes fiber using the setup described above. The maximum efficiency measured at the diffraction limit for both fibers is lower than that theoretically predicted in Fig.~\ref{fig:eta_fno_DL} by $\sim 25\%$, for which we identify three causes: 1. Coupling in the testbed is done at \textsf{\textflorin}$/5.56$, the optimum for the SMF but slightly faster than the \textsf{\textflorin}$/6.8$ required to maximally couple into a $6$-mode fiber. 2. The aberrations in the optical train amount to an rms wavefront error of $80 - 120$ nm depending on the telescope in place and can only be flattened down to a minimum of $20$ nm rms error by closing the loop. The reduction of the theoretical limit (at an open loop) due to aberrations was confirmed by modelling the trains in Zemax where a physical optics module can calculate the overlap between the fundamental mode and the deformed PSF. Alignment was improved until those theoretical limits were reached. 3. Insertion and transmission losses in the fibers introduce a small reduction in the measured efficiencies as compared to the other factors. 

\begin{figure}

	\includegraphics[width=0.9\columnwidth]{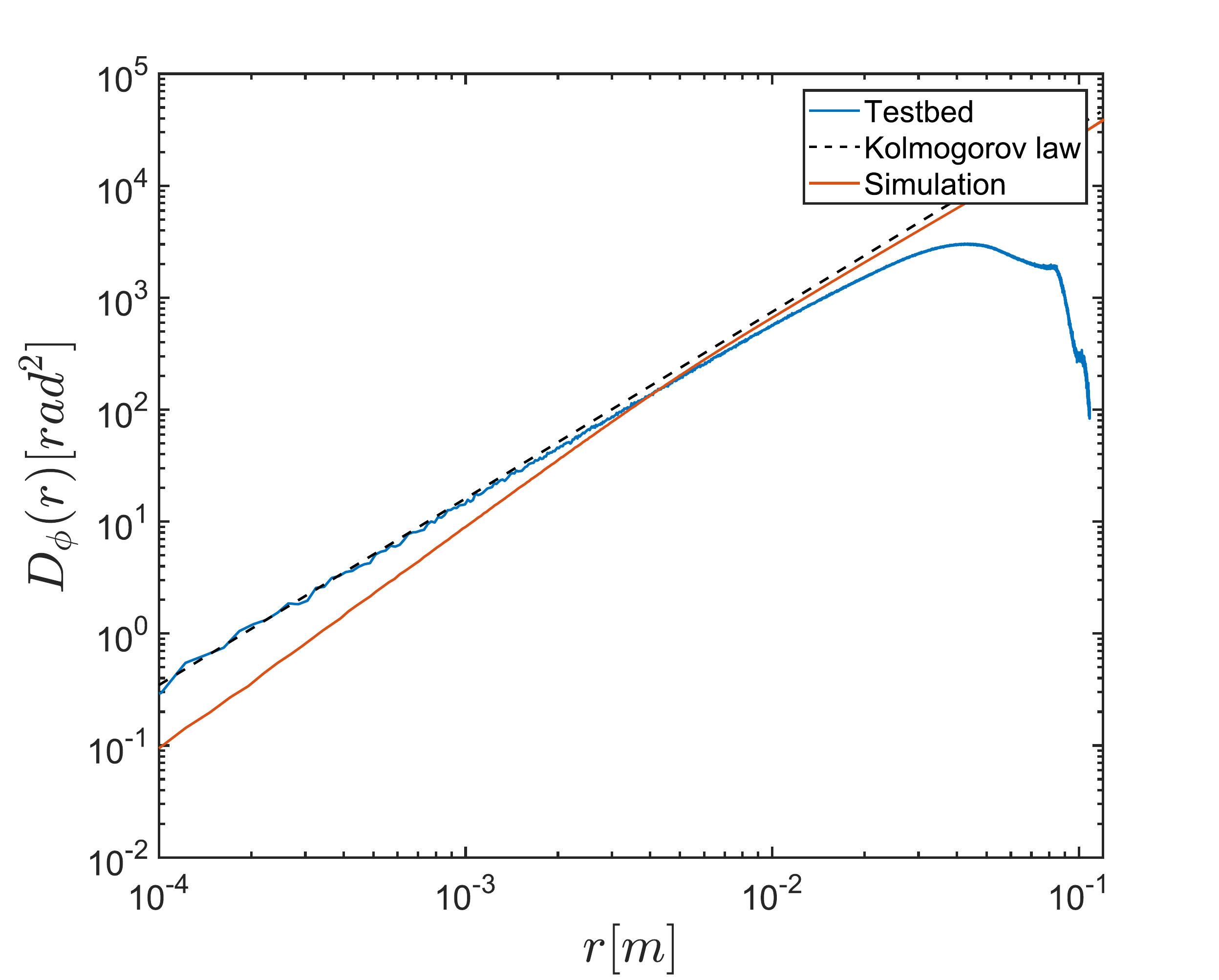}
    \caption{Structure functions of the testbed phase screen (blue line) and the model realizations (orange line). Plotted also is Kolmogorov's $5/3$ law of turbulence (dashed line).}
    \label{fig:SF}
\end{figure}

The reasons mentioned for the mismatch above cancel out for the gain curves and therefore we choose to qualify the models using them. 
At higher $D/r_0$ values the theoretical gain remains constant but a drop is seen in the experimental values for $D/r_0 > 16$. This is a result of filtering the high-order eigenmodes of the DM to calculate the command matrix in Eq.~\eqref{eq:AO} when closing the loop. The high-order eigenmodes have smaller eigenvalues and therefore cause the actuator commands to saturate and the loop to diverge when high spatial frequency shapes are requested of the DM. Only $3$ eigenmodes are filtered for the cases where $D/r_0 < 10$ but more are filtered dynamically by the control code as $D/r_0$ is increased and the loop starts to diverge.  
Moreover, the phase screen of the testbed was calculated as a Fourier series (FS) expansion over an area equal to the size of the screen while the phase screen realizations for the model were calculated over an area $40\times$ larger to minimize the inherited periodicity in the FS-generated phase. The testbed screen therefore has a structure function (Eq.~\eqref{eq:SF}) that deviates from Kolmogorov's $5/3$ law at greater separations while the model's screens adhere to the law more closely. The underrepresented low frequencies in the testbed screen cause a higher coupling efficiency into the fibers for the uncorrected case while the AO-corrected case remain unaffected (c.f. Fig.~\ref{fig:SF}). Furthermore, a large ensemble of unique screens cannot be achieved for larger beam diameters due to the finite size of the phase screen.

\begin{figure}

	\includegraphics[width=0.9\columnwidth]{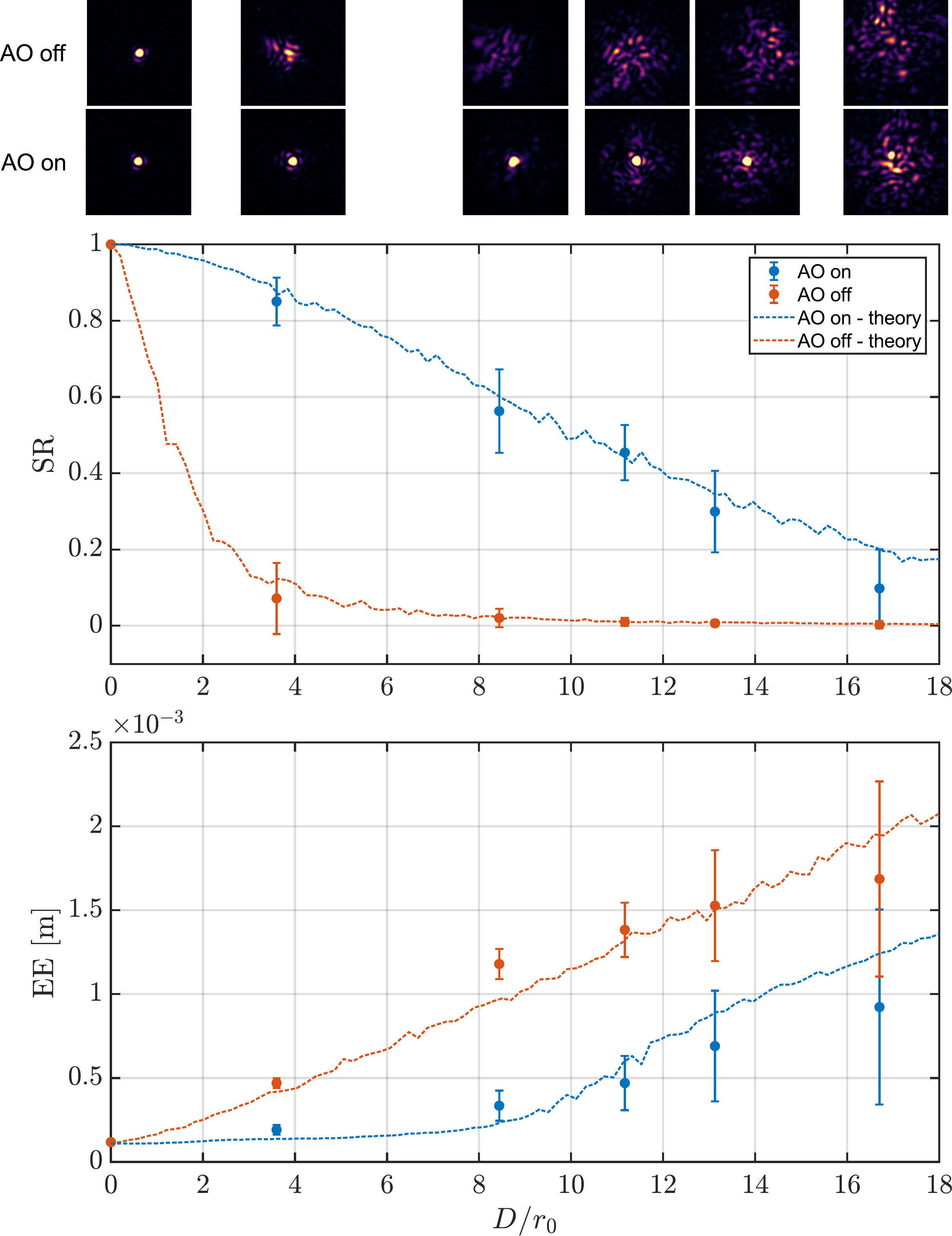}
    \caption{Top: Short exposure PSF images taken with the loop open and then closed for increasing $D/r_0$. Middle: SR dependence on $D/r_0$ for the corrected and the uncorrected cases. Bottom: PSF 50\% encircled energy at \textsf{\textflorin}$/74$ dependence on $D/r_0$. Simulation results from the theoretical model are also plotted for reference.}
    \label{fig:SR}
\end{figure}

Fig. \ref{fig:SR} shows the drop in the Strehl ratio (SR) and the increase in the encircled energy (EE) as $D/r_0$ increases. The resemblance between the $\eta$ curves for the SMF in Fig. \ref{fig:eta_dr0} (when normalized to have $\eta = 1$ at the diffraction limit) and the SR curves in Fig. \ref{fig:SR} is a result of the SMF being nearly a point-like sampler of the center of the PSF as first noticed by \citet{v_coude_2000}. The knee in the EE curve for the AO corrected case at $D/r_0\sim9$ indicates the transition between the total and the partial correction regimes where $r_0$ projected on the DM becomes smaller than the inter-actuator spacing. The PSF is no longer contained by the LOAO system and starts to broaden at a rate equal to that of the uncorrected case. Strehl ratio drops linearly throughout both regimes for the corrected wavefronts since a central core is always present in the PSF for $D/r_0 < 18$ as shown in the images at the top of Fig. \ref{fig:SR}. The deviation of the data points at $D/r_0 = 16.7$ from the theoretical prediction is again due to the periodicity of the phase screen as discussed above.    

\section{Discussion}
\label{sec:discuss}
The results for the effect of the \textsf{\textflorin}$/\#$ of the telescope on coupling in Fig.~\ref{fig:eta_fno_DL} match those reported by~\citet{horton} for the diffraction-limited case. The extension into the seeing limit and the study here of how the optimum geometry varies with turbulence was necessary to confirm that diffraction-limited values remain valid for all cases. FMFs coupling curves (see Fig.~\ref{fig:eta_dr0}) for the uncorrected case can be cross-checked against a number of references that studied coupling for FSO communication.~\citet{tedder20} measured coupling for a $15$-modes fiber up to $D/r_0 = 8.6$ and the $6$ dB loss reported is comparable to the $0.22$ coupling efficiency for the $19$ modes fiber we calculated.~\citet{zheng} simulated and measured received power into SMFs and their results can be directly compared to the drop in efficiency and increase of the standard deviation with $D/r_0$. For example, the $1.25\%$ efficiency they reported for $D/r_0 = 10$ matches the calculated $\eta = 1.24\%$ in Fig.~\ref{fig:eta_dr0}. The addition here is the inclusion of AO correction and the focus on the modal counts most relevant to tapered photonic lanterns.

With an experimentally-verified model like the pipeline detailed in this work, one has the tool to decide if a multiplexed photonic instrument fed by an LOAO-assisted photonic lantern provides an advantage over a single device in terms of SNR. A typical scenario is a telescope equipped with an AO system that has a maximum number of modes and a temporal bandwidth it can correct for but can also function at decreased capabilities in the absence of a bright enough guide star. A photonic lantern can then be appropriately sized given the source brightness, the available exposure time and the detector specifications. The availability of photonic components (lantern and IO devices) in different sizes and quantities for a quick exchange should be feasible although packaging and alignment techniques need to be perfected to minimize losses and downtime.

Turnkey, general-purpose AO systems are now available from a variety of vendors. The use of pre-engineered hardware and software make such solutions affordable ($\sim\$10^5$) for midsize telescopes where astrophotonic technologies could be employed first. This is in contrast to the custom-made, large-scale AO projects currently existing or being considered for the very and extremely large telescopes. FSO communication applications in particular are set to benefit from the anticipated ubiquity of low-cost AO systems~\citep{leonhard2016}.     


Once a verified model for coupling through turbulence in the presence of AO compensation is available, color maps like those in Fig.~\ref{fig:SNRs} can be generated for any given celestial target and scientific camera. A quick multi parameter scan shows the regime ($D/r_0$, $N_{\mathit{ph}}$, and $\sigma_{\mathit{RON}}$) under which multiplexing becomes beneficial and the optimum number of channels that one should opt for to get the most cost-efficient, i.e. least number of channels, that maximizes the SNR above a certain threshold.  


The study done here concerns H-band astronomy as this is the band where the operating ranges of current AO and photonic technologies overlap. Results from simulations and the experiment detailed above are for $\lambda = 1550$ nm but a recalculation at a different wavelength between $1500$ and $1800$~nm is possible without modification as, in principle, the physics and the assumptions made remain valid. In general, Fried parameter is smaller towards the blue $r_0 \propto \lambda^{6/5}$ (c.f. Eq. \eqref{eq:r0}) effectively squeezing the curves in Fig.~\ref{fig:eta_dr0} to the left as the wavelength decreases. however the number of modes supported by a given wave\-guide increases approximately as $\sim \lambda ^{-2}$, hence increasing the total coupling efficiency. The general case of an unobscured circular aperture was considered for the simulations and the experiments reported here but obscurations, spiders and segmented pupils can be readily included in the models when necessary. An on-sky test of fully packaged photonic lanterns of different sizes assisted by an LOAO shall complement this work.




\section{Conclusions}
\label{sec:conclusions}


We presented a feasibility study and made the case for H-band multiplexed astrophotonic instruments fed by AO-assisted photonic lanterns. A numerical simulation was completed to find the compromise between the complexity of the AO system and the size of the photonic lantern to maximize sensitivity. Photonic lanterns sustain few modes at their multimode ports and the optimum f-numbers for coupling into FMFs deviate from the geometric predictions as Fig.~\ref{fig:eta_fno_DL}  depicts. 
An LOAO system can boost the coupling of atmospherically-distorted starlight into FMFs and photonic lanterns manyfold ($2 \sim 100$, c.f. Fig.~\ref{fig:eta_dr0}).  
The SNR of an accumulated signal detected at the output of a multiplexed instrument depends on the photons flux that can be coupled into the instrument, the number of channels over which they are split by a photonic lantern and the RON of the detector used (c.f. Fig.~\ref{fig:SNRs}). 
The realm where such an approach offers an advantage over singular standalone devices fed by SMFs is therefore defined by the aperture size, the science target, and the detector capability. 

With the prevalence of LOAO systems, the continuing improvement of low noise infrared detectors, and the imminent adoption of photonic technologies by astronomy, multiplexed photonic instruments will soon become advantageous for midsize and large telescopes. Immediate applications for extremely large telescopes could be considered for  AO-supported multi-object spectrographs that are currently being designed for the next generation of ELTs, such as MOSAIC for the ELT \citep{jagourel_mosaic_2018, morris_phase_2018}.

\section*{Acknowledgements}

The author is most thankful to Stefano Minardi for his valuable guidance during the early stages of this work. The author is also thankful to Peter Tuthill for suggestions on applying mode-selective photonic lanterns, to Sergio Leon-Saval for comments on scrambling and the geometric considerations of lanterns design, to Alan G\"unther for help with thermal effects in the lab, and to the anonymous reviewer for useful remarks about modal noise in fibers.

The authors acknowledge financial support from the Federal Ministry of Education and Research (BMBF) under grant number 03Z22AN11.

\section*{Data Availability}
 	
The data underlying this article will be shared on reasonable request to the corresponding author.




\bibliographystyle{mnras}
\bibliography{mnras_template} 

\bsp	
\label{lastpage}
\end{document}